\documentclass[useAMS,usenatbib]{mn2e}
\usepackage{graphicx,psfig,cite,epsf}  
\usepackage{times}
\usepackage{subfigure}
\usepackage{textcomp}
\usepackage{multirow}

\def\apjl{ApJ}
\def\apjl{ApJ}                
\def\apjs{ApJS}               
\def\ao{Appl.~Opt.}           
\def\aap{A\&A}                

\title[Testing Rate Dependent corrections on GX 13+1]{Testing Rate Dependent corrections on timing mode EPIC-pn spectra of the accreting Neutron Star GX 13+1}
\author[Pintore et al.]{\Large{F. Pintore$^1$, A. Sanna$^1$, T. di Salvo$^3$, M. Guainazzi$^2$, A. D'A\`i$^1$, A. Riggio$^1$, L. Burderi$^1$, R. Iaria$^3$, N. R. Robba$^3$}\\
$^1$ Universit\`a degli Studi di Cagliari, Dipartimento di Fisica, SP Monserrato-Sestu, KM 0.7, 09042 Monserrato, Italy \\
$^2$ European Space Astronomy Center of ESA, Apartado 50727, 28080 Madrid, Spain\\
$^3$ Dipartimento di Fisica e Chimica, Universit\'a di Palermo, via Archirafi 36 - 90123 Palermo, Italy}

\begin{document}

\maketitle

\begin{abstract}
When the EPIC-pn instrument on board \textit{XMM-Newton} is operated in Timing mode, high count rates ($>100$ cts s$^{-1}$) of bright sources may affect the calibration of the energy scale, resulting in a modification of the real spectral shape. The 
corrections related to this effect are then strongly important in the study of the spectral properties. Tests of these calibrations are more suitable in sources which spectra are characterised by a large number of discrete features. Therefore, in this work, we carried out a spectral analysis of the accreting Neutron Star GX 13+1, which is a dipping source with several narrow absorption lines and a broad emission line in its spectrum.
We tested two different correction approaches on an \textit{XMM-Newton} EPIC-pn observation taken in Timing mode: the standard Rate Dependent CTI (RDCTI or \textit{epfast}) and the new, Rate Dependent Pulse Height Amplitude (RDPHA) corrections. 
We found that, in general, the two corrections marginally affect the properties of the overall broadband continuum, while hints of differences in the broad emission line spectral shape are seen. On the other hand, they are dramatically important for the centroid energy of the absorption lines. In particular, the RDPHA corrections provide a better estimate of the spectral properties of these features than the RDCTI corrections. Indeed the discrete features observed in the data, applying the former method, are physically more consistent with those already found in other \textit{Chandra} and \textit{XMM-Newton} observations of GX 13+1.
\end{abstract}


\section{Introduction}

The calibration of the detectors used in astronomy is fundamental for the sake of scientific studies. However, the processes at the basis of the calibrations are often complex especially for instruments which are onboard satellites. Indeed their calibrations have to quickly evolve with time as the space environment does not favour the stability of the instruments, for example due to the impact with micrometeorites, excessive irradiation by high energy particles or unexpected effects that were not observed on the ground. Focusing our attention to X-ray satellites, they are also highly affected by fluorescence emission lines produced by material in the satellite environment when irradiated by X-ray photons. Furthermore, the calibration of the instruments are also carried out using an X-ray source onboard of the satellite which produces only a few emission lines and does not allow a precise calibration of the whole energy range. 

Here we focus our attention on the impact of the calibration of the energy scale in the EPIC-pn instrument \citep{struder01} on-board XMM-Newton \citep{jansen01}, when this instrument is operated in Timing Mode and observing bright sources ($>100$ count s$^{-1}$). It has been noticed that the large amount of photons, i.e. energy, which these bright sources deposit on the CCD may affect the calibration of the energy scale. In particular, it distorts the observed spectral shape of the sources, altering the scientific results. In order to account for this effect, two approaches were developed: one is calibrated on the spectrum in Pulse Invariant (PI), assuming an astrophysical model of the spectrum in the 1.5-3 keV energy band; while the other one acts on the Pulse Height Analyser information (PHA, Guainazzi 2013). The former, called Rate Dependent Charge Transfer inefficiency (RDCTI or \textit{epfast}; Guainazzi et al. 2008, XMM-CAL-SRN-248\footnote{http://xmm2.esac.esa.int/docs/documents/CAL-SRN-0248-1-0.ps.gz}), was historically developed to correct for Charge Transfer Inefficiency (CTI) and X-ray loading (XRL), although their energy-dependence is based on unverified assumptions (Guainazzi $\&$ Smith 2013, XMM-CAL-SNR-0302\footnote{http://xmm2.esac.esa.int/docs/documents/CAL-SRN-0302-1-5.pdf}). Instead in the second approach, called Rate Dependent Pulse Height Amplitude (RDPHA), the energy scale is calibrated by fitting the peaks in derivative PHA spectra corresponding to the Si-K ($\sim1.7$ keV) and Au-M ($\sim2.3$ keV) edges of the instrumental response, where the gradient of the effective area is the largest. It also includes an empirical calibration at the energy of the transitions of the $K_{\alpha}$ Iron line (6.4-7.0 keV; Guainazzi, 2014, XMM-CAL-SRN-0312\footnote{http://xmm2.esac.esa.int/docs/documents/CAL-SRN-0312-1-4.pdf}). The RDPHA approach avoids any assumption on the model dependency in the spectral range around the edges and it is also calibrated in PHA space before events are corrected for gain and CTI. 
In order to test the goodness of these corrections, bright sources with a number of features (in absorption or emission) are needed.

To investigate these two approaches, we selected the bright source GX 13+1 as a test-study. GX 13+1 is a low mass X-ray binary (LMXB) source which is a well known persistent accreting neutron star (NS) at the distance of $7\pm1$ kpc. Its companion is an evolved K IV mass-donor giant star \citep{bandyopadhyay99}. In particular, GX 13+1 is a dipping source \citep{corbet10,diaz12} which has probably shown periodic dips due to the orbital motion (\citealt{iaria14}) during the last decades. It was suggested that dips are more likely produced by optically thick material at the outer edge of the disc created by the collision between the accretion flow from the companion star and the outer disc (e.g. \citealt{white82}) or by outflows in the outer disc. The orbital period of GX 13+1 is 24.52 days, making this source the second LMXB with the longest orbital period after GRS 1915+105.

The continuum emission of GX 13+1 can be described with the combination of a multicolour blackbody plus a cold, optically thick comptonisation component (e.g. \citealt{homan04,diaz12}). Because the comptonisation is optically thick, it can be also approximated by a blackbody component, lightening the fit calculation given that the properties of the comptonising component (electron temperature and optical depth) are usually poorly constrained \citep{ueda01,sidoli02,ueda04}. The spectra show also the existence of several spectral features \citep{diaz12,dai14}. These are associated with a warm absorbing medium close to the source and produced by outflows from the outer regions of the accretion disc. The warm absorber is present during all the orbital period and might become denser during the dips episodes, suggesting a cylindrical distribution around the source. It also results more opaque, and probably cloudy, close to the plane of the disc. In addition, the absorption lines associated to the warm absorber are produced by highly ionised species and indicate bulk outflow velocities of $\sim 400$ km s$^{-1}$ \citep{ueda04,madej13,dai14}.

Furthermore, a broad, emission component at the energy of the K-shell of the Iron XXI-XXVI was found and interpreted as reflection of hard photons from the surface of the accretion disc \citep{diaz12}. The broadness of the line has been suggested not to be produced by relativistic effects but by Compton broadening in the corona, although this interpretation has been questioned \citep{cackett13}.

GX 13+1 is therefore an ideal candidate to test the RDCTI and RDPHA calibration approaches as it shows a simple continuum characterised by a number of narrow absorption features and a broad emission line.

\section{Data Reduction}
\label{data_reduction}

We carried out a spectral analysis on one \textit{XMM-Newton} observation (Obs.ID. 0122340901) taken in Timing mode, of the accreting NS GX 13+1. Data were reduced using the latest calibrations (at the date of April 25th, 2014) and Science Analysis Software (SAS) v. 13.5.0. At high count rates, X-ray loading and CTI effects have to be taken into account as they affect the spectral shape and, in particular, they produce an energy shift on the spectral features. 
In order to optimize the data reduction, we generated two EPIC-pn events files, each one created according to the RDCTI and RDPHA corrections: for the RDCTI corrections, we made use of the standard \textit{epfast} corrections\footnote{http://xmm.esac.esa.int/sas/current/documentation/threads/EPIC$\_$ reprocessing.shtml} adopting the following command: ``{\sc epproc runepreject=yes withxrlcorrection=yes runepfast=yes}''; for the RDPHA corrections, we reprocessed the data using the command: ``{\sc epproc runepreject=yes withxrlcorrection=yes runepfast=no withrdpha=yes}''. The command ``{\sc runepfast=no withrdpha=yes}'' has to be explicitly applied in order to avoid the combined use of both corrections. We note that the adopted RDCTI (\textit{epfast}) task is the latest released version and its effect on data is dissimilar from versions of \textit{epfast} older than May 23$^{rd}$, 2012. Indeed the older versions combined XRL and rate-dependent CTI corrections in a single correction, while they are now applied separately, each with its appropriate calibration. Hence the calibrations due to different \textit{epfast} versions might provide different results.

For each EPIC-pn event file, we extracted the spectra from events with {\sc pattern}$\leq 4$ (which allows for single and double pixel events) and we set `{\sc flag}=0' retaining events optimally calibrated for spectral analysis. Source and background spectra were then extracted selecting the ranges RAWX=[31:41] and RAWX=[3:5], respectively.
We generated the auxiliary files using \textit{arfgen} and setting ``detmaptype=psf'' and  ``psfmodel=EXTENDED'', using the calibration file ``XRT3$\_$XPSF$\_$0016.CCF'' (Guainazzi et al., 2014; XMM-CAL-SRN-0313\footnote{http://xmm2.esac.esa.int/docs/documents/CAL-SRN-0313-1-3.pdf}). EPIC-pn spectra were subsequently rebinned with an oversample of 3 using \textit{specgroup}.

Finally, all RGS spectra were extracted using the standard \textit{rgsproc} task, filtered for periods of high background and grouped with a minimum of 25 counts per noticed channel.

The RGS and EPIC-pn spectra were then fitted simultaneously using {\sc xspec} V. 12.8.1 \citep{arnaud96}, in the range 0.6-2.0 keV and 2.0-10.0 keV, respectively. 

We also compared the EPIC-pn data to a \textit{Chandra} observation, with Obs.ID 11814. In particular, we analysed the data of the High Energy Grating (HEG) instrument onboard \textit{Chandra}. Since the data reduction and extraction process is described in \citet{dai14}, we suggest the reader to reference that paper for more details.

\subsection{Pile-up}

\begin{figure}
\center
\hspace{-0.63cm} \includegraphics[height=9.0cm,width=7.0cm,angle=270]{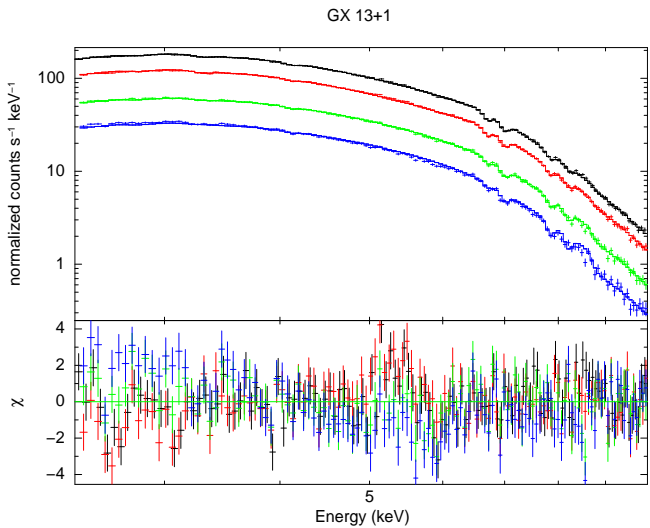}
\caption{EPIC-pn spectra where any (\textit{black}), one (\textit{red}), three (\textit{green}) and five (\textit{blue}) columns were excised in order to test for pile-up effects. We also show the residuals obtained adopting the model {\sc phabs*edge*(nthcomp + gauss)} and fitting the spectra simultaneously. We found consistency between the spectra obtained when removing three and five columns of the CCD.}
\label{comparison_pile_up}
\end{figure}

The source has a mean count rate in the EPIC-pn detector of $\sim700$ cts s$^{-1}$, close to the nominal threshold for pile-up effects, that is $>800$ cts s$^{-1}$ in Timing mode. In order to test the presence of pile-up effects in the EPIC data, we initially made use of the SAS tool \textit{epatplot}. It provides indications that the data are affected by pile-up which can be widely corrected excising the three brightest central column (RAWX=[35:37]).
However, the best test can be done comparing the residuals of the spectra with none, one, three and five columns excised, where a fit with a same spectral model is adopted. Hence we selected the range 2.4-10 keV of the four EPIC-pn spectra (RDPHA corrected) and we fitted them simultaneously with an absorbed {\sc nthcomp} model (\citealt{zdiarski96}), letting only the normalizations between the spectra free to vary. We also added an absorption edge and Gaussian models to take into account some absorption narrow lines and a broad emission line (see next sections for major details). We obtained a reduced $\chi^2$ of 1.9 ($\chi^2=874.93$ for 451 degrees of freedom). We show the best fit and its residuals in Figure~\ref{comparison_pile_up}.
Inspecting the residuals, spectra extracted after excising three and five columns are consistent, confirming that removing only three CCD column corrects for pile-up effects. Therefore in the spectral analyses of the next sections we will only consider the spectrum extracted removing the three central column.
We also highlight that the same findings hold also for the RDCTI corrected spectra.

\section{Spectral analysis}
\label{analysis}
We analysed spectra extracted from RDPHA and RDCTI corrected event files and we adopted the same continuum for both of them. The neutral absorption is described with the {\sc phabs} model, using the abundance of \citet{andres89}. The continuum is instead based on a blackbody component ({\sc bbody} in {\sc xspec}) plus a comptonisation component ({\sc nthcomp} in {\sc xspec}; \citealt{zdiarski96}).
We note that leaving the seed photons temperature (kT$_{seed}$) free to vary makes this parameter totally unconstrained; therefore we linked it with the blackbody temperature, assuming that the seed photons for comptonisation are provided by the inner regions of the accretion disc.  
 
\begin{figure}
\center
\hspace{-0.63cm} \includegraphics[height=9.0cm,width=7.0cm,angle=270]{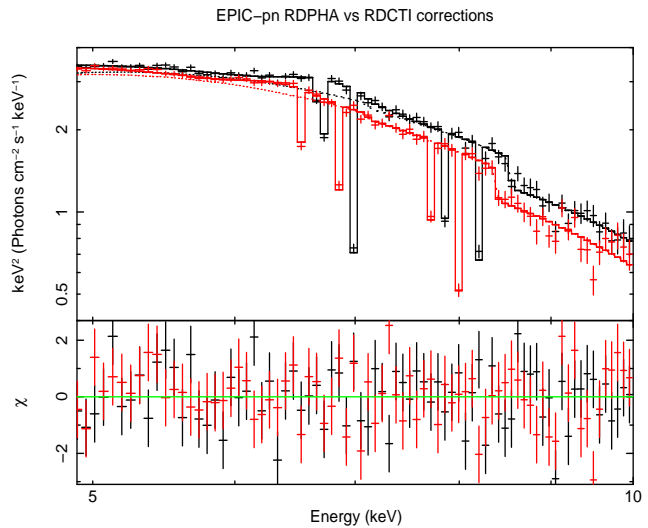}
\caption{ Comparison of the unfolded ($Ef(E)$) EPIC-pn spectra in the 5-10 keV energy band, applying RDPHA (\textit{black} spectrum) and RDCTI (\textit{red} spectrum) corrections. Both dataset are fitted with their corresponding best fit models, {\sc edge$\cdot$phabs$\cdot$(bbody+nthcomp+6 gaussian)} showed in Table~\ref{table_continuum_gauss_rdpha} and ~\ref{gauss_rdpha} (the emission feature is taken into account). Below 5 keV, the spectra are generally consistent. Instead, above this threshold, clear discrepancies in the features and the continua are seen (see text).}
\label{comparison_rdpha_rdcti}
\end{figure}

We introduced a multiplicative constant model in each fit in order to take into account the diverse calibration of RGS and EPIC instruments. We fixed to 1 the constant for the EPIC-pn spectrum and allowed the RGS constants to vary. In general, this parameter range does not vary more than 10$\%$ in comparison with the EPIC-pn constant.

\begin{table*}
\footnotesize
\begin{center}
\caption{Best fit spectral parameters obtained with the absorbed {\sc bbody+nthcomp} model plus {\sc diskline} or {sc reflionx}.The Gaussian lines and the absorption edge are always taken into account. Errors are at 90$\%$ for each parameter.}
\label{table_continuum_gauss_rdpha}
\scalebox{0.85}{\begin{minipage}{18.0cm}
\begin{tabular}{lllllllll}
\hline
Model & Component & \multicolumn{6}{c}{1 col. removed$^1$} \\ 
\\
\multicolumn{1}{c}{(1)} & \multicolumn{1}{c}{(2)} & \multicolumn{1}{c}{(3)} & \multicolumn{1}{c}{(4)} & \multicolumn{1}{c}{(5)} & \multicolumn{1}{c}{(6)} & \multicolumn{2}{c}{(7)}  & \multicolumn{1}{c}{(8)} \\
\\
& & RDPHA & RDCTI & RDPHA & RDCTI & \multicolumn{2}{c}{RDPHA$^{**}$} & RDCTI \\
\\
{\sc phabs} &N$_H$ (10$^{22}$ cm$^{-2}$)$^a$ &$2.67^{+0.05}_{-0.05}$ & $2.70^{+0.05}_{-0.05}$ & $2.68^{+0.06}_{-0.05}$ & $2.68^{+0.05}_{-0.05}$& $2.90^{+0.09}_{-0.06}$ &$3.66^{+0.06}_{-0.09}$ & $3.4^{+0.1}_{-0.1}$\\
\\
{\sc bbody} &kT$_{bb}$ (keV)$^b$& $0.55^{+0.02}_{-0.02}$ & $0.52^{+0.02}_{-0.02}$ & $0.53^{+0.02}_{-0.03}$ & $0.54^{+0.02}_{-0.02}$& $0.53^{+0.1}_{-0.2}$ & - &$0.33^{+0.03}_{-0.02}$\\
& Norm.$^c$ & $0.036^{+0.001}_{-0.01}$ & $0.037^{+0.001}_{-0.01}$ & $0.034^{+0.001}_{-0.002}$ & $0.037^{+0.001}_{-0.01}$& $0.002^{+0.003}_{-0.002}$& - &$0.022^{+0.009}_{-0.009}$\\
\\
{\sc nthcomp} & kT$_{e}$ (keV)$^d$&$1.21^{+0.04}_{-0.01}$& $1.16^{+0.01}_{-0.01}$  & $1.18^{+0.01}_{-0.03}$ &$1.17^{+0.02}_{-0.04}$& $1.20^{+0.01}_{-0.04}$& $1.20^{+0.02}_{-0.05}$ & $1.13^{+0.01}_{-0.01}$\\
&$\Gamma$$^e$&$1.0^{+0.2}_{-*}$& $1.0^{+0.2}_{-*}$ & $1.0^{+0.1}_{-*}$ & $1.0^{+0.2}_{-*}$& $1.44^{+0.1}_{-0.2}$& $1.40^{+0.3}_{-*}$ & $1.4^{+0.1}_{-*}$\\
&kT$_{0}$$^f$& = kT$_{bb}$& = kT$_{bb}$ & = kT$_{bb}$ & = kT$_{bb}$& = kT$_{bb}$& $1.2^{+0.6}_{-0.4}$ & = kT$_{bb}$\\
\\
{\sc diskline} & Energy (keV)$^g$& - & -&6.63$_{-0.03}^{+0.06}$ & 6.60$_{-0.03}^{+0.03}$&- & -& -\\
& Betor10 $^h$ & - & - &-2.37$_{-0.1}^{+0.09}$ & -2.43$_{-0.1}^{+0.2}$& -& -& -\\
& R$_{in}$ (R$_g$)$^i$ & -& -&16$_{-10}^{+7}$ & 6$_{-*}^{+2}$ & -& -& -\\
& Inclination (degree)$^l$ & -& - &90$_{-37}^{*}$ & $<21$& -& -& -\\
& Norm.$^m$ & -& - &1.1$_{-0.3}^{+0.4}$$\times10^{-2}$ & 6.91$_{-0.2}^{+0.09}$$\times10^{-3}$& -& -& -\\
\\
{\sc highecut} &cutoffE  (keV)$^n$& -&-&-&-& \multicolumn{3}{c}{0.1 (frozen)}\\
 &foldfE (keV)$^o$&- &- &-&-& \multicolumn{3}{c}{2.7$\times$kT$_{e}$ (frozen)}\\
\\
{\sc rdblur} & Betor10 $^h$& - & -& -& -& -2.82$_{-0.09}^{+0.1}$& -2.82$_{-0.4}^{+0.3}$&-2.57$_{-0.1}^{+0.1}$\\
 & R$_{in}$ (R$_g$)$^i$ & - & -& -& -& 7.0$_{-1}^{+2}$& 10$_{-4}^{+5}$&15$_{-6}^{+4}$\\
 & Inclination (degree)$^l$& - & -& -& -&66$_{-3}^{+6}$& 52$_{-5}^{+8}$&$>60$\\
\\
{\sc reflionx} & Fe/solar$^p$ & - & -& -& -& 3.0$_{-1.3}^{+*}$& 3.0$_{-0.5}^{+*}$&3.0$_{-1.0}^{+*}$\\
 & $\xi$ (erg cm s$^{-1}$)$^q$& - & -& -& -& $<20$& $1060_{-86}^{+73}$&220$_{-10}^{+17}$\\
\hline
&$\chi^2/dof$$^r$&904.01/836& 891.08/836&900.26/834& 885.94/834&923.56/835& 929.42/836&922.98/835\\
\end{tabular} 
\end{minipage}}
\end{center}
\begin{flushleft} $^1$ EPIC-pn spectrum corrected for pile-up removing the central, brightest column; $^a$ Column density; $^b$ Blackbody temperature and seed photons temperature of the {\sc nthcomp}; $^c$ Normalization of the {\sc bbody} component in unity of $L_{39}^2/D_{10kpc}$, where $L_{39}$ is the luminosity in unity of 10$^{39}$ erg s$^{-1}$) and $D_{10kpc}$ is the distance in unity of 10 kpc; $d$ Electrons temperature of the corona; $^e$ Photon index; $^f$ Seed photons temperature (usually equal to kT$bb$); 
$^g$ Energy of the relativistic line; $^h$ Power law dependence of emissivity; $^i$ Inner radius in terms of gravitational radius R$_g$; $^l$ Inclination angle of the binary system; $^m$; Normalization of the model in {\sc xspec} unit; $^n$ low energy cut-off of the reflection component; $^o$ high energy cut-off of the reflection component, set as 2.7 times the electron temperature of the comptonisation model; $^p$ Ratio of Iron and hydrogen abundance; $^q$ ionisation parameter; $^r$ reduced $\chi^2$ of the best fit including the absorption/emission features shown in Table~\ref{gauss_rdpha}. \\
$*$: the value pegged at its higher/lower limit; $^{**}$ Two alternative models (with and without {\sc bbody} component) are shown for this spectrum. \end{flushleft}
\end{table*}

\begin{table}
\footnotesize
\begin{center}
\caption{Best fitting absorption features evaluated adopting a Gaussian or {\sc xstar} model and introducing an absorption edge. Errors are at 90$\%$ for each parameter.}
\label{gauss_rdpha}
\scalebox{0.8}{\begin{minipage}{9.0cm}
\begin{tabular}{llll}
Model & Component & \multicolumn{2}{c}{1 col. removed$^1$} \\ 
\\
& & RDPHA & RDCTI\\
\\
\hline
\multicolumn{4}{c}{\sc emission feature}\\
\\
\textit{\sc Gaussian} & E$_{line}$ (keV)$^a$&$6.69_{-0.07}^{+0.07}$ & $6.33_{-0.09}^{+0.08}$\\
& $\sigma_v$ (keV)$^b$&$0.34_{-0.08}^{+0.1}$ & $0.24_{-0.08}^{+0.1}$\\
& Norm.$^c$ &$5.9^{+0.2}_{-0.2} \times10^{-3}$ & $1.2^{+0.1}_{-0.1} \times10^{-2}$\\
\hline
\multicolumn{4}{c}{\sc absorption features}\\
\\
\textit{\sc Gaussian$^*$} & E$_{line}$ (keV)&$2.26_{-0.02}^{+0.02}$ & $2.22_{-0.08}^{+0.08}$  \\
& Norm. &$6.2^{+0.2}_{-0.2} \times10^{-3}$ & $1.6^{+0.2}_{-0.2} \times10^{-3}$\\
\\
\textit{\sc Gaussian} & E$_{line}$ (keV)&$6.70_{-0.02}^{+0.02}$ & $6.55_{-0.02}^{+0.02}$  \\
& Norm. &$1.9^{+0.4}_{-0.7} \times10^{-3}$ & $1.6^{+0.7}_{-0.4} \times10^{-3}$\\
\\
\textit{\sc Gaussian} & E$_{line}$ (keV)&$6.99_{-0.01}^{+0.01}$ & $6.83_{-0.02}^{+0.02}$\\
& Norm. &$2.6^{+0.4}_{-0.4} \times10^{-3}$ & $1.8^{+0.6}_{-0.5} \times10^{-3}$\\
\\
\textit{\sc Gaussian} & E$_{line}$ (keV)&$7.86_{-0.03}^{+0.04}$ & $7.69_{-0.04}^{+0.04}$\\
& Norm. &$1.1^{+0.3}_{-0.3} \times10^{-3}$ & $1.1^{+0.3}_{-0.3} \times10^{-3}$\\
\\
\textit{\sc Gaussian} & E$_{line}$ (keV)&$8.19_{-0.05}^{+0.05}$ &$7.99_{-0.03}^{+0.1}$\\
& Norm. &$9^{+3}_{-3} \times10^{-4}$ &$1.3^{+0.3}_{-0.3} \times10^{-3}$\\
\\
\textit{\sc edge} & E$_{edge}$ (keV)&$8.7_{-0.1}^{+0.2}$ & $8.38_{-0.08}^{+0.08}$\\
 & $\tau$ &$0.13_{-0.05}^{+0.06}$ & $0.27_{-0.05}^{+0.05}$\\
\hline
\hline
\multirow{3}{*}{\sc xstar} & N$_H^{abs}$ (10$^{22}$ cm$^{-2}$)$^d$ & $60_{-30}^{+20}$&$4_{-1}^{+1}$  \\
& Log($\xi_{abs}$)$^e$& $4.2\pm0.1$ &  $3.41\pm0.09$\\
& z$^{abs}$ (km s$^{-1}$)$^f$& $-330_{+500}^{-260}$ & $-2600_{+1300}^{-1800}$ \\
\hline
\\
\end{tabular} 
\end{minipage}}
\end{center}
$^1$  EPIC-pn spectrum corrected for pile-up removing the central, brightest column; $^a$ Energy of the feature;  $^b$ Line width in keV $^c$ Normalization of the feature ({\sc xspec} units); $^d$ columns density of the warm absorber; $^e$ ionisation parameter of the warm absorber; $^f$ blueshift velocity of the warm absorber. $*$ This line can be associated to a residual of calibration around the instrumental Au-M edge. \\
\end{table}

\subsection{Broadband continuum and narrow absorption features}
\label{absfeatures}

In Table~\ref{table_continuum_gauss_rdpha} (columns $\#3$ and 4), we show the best fit parameters obtained with the continuum model described in the previous section. 
RDPHA and RDCTI corrections give similar continuum parameters: they are both well described by a {\sc bbody} with a temperature of $\sim 0.55$ keV and the comptonisation component shows a marked roll-over into the \textit{XMM-Newton} bandpass. Indeed, the electron temperature (kT$_e$) is consistent with $\sim1.2$ keV, while the powerlaw photon index ($\Gamma$) lies at 1.0, although it pegged to the lower limit. 

Not surprisingly, several features (in absorption and emission) are clearly observed in the EPIC-pn spectra and we initially model all of them with Gaussian components, following \citet{diaz12} and \citet{dai14}. For the absorption lines, we fixed at zero the dispersion width ($\sigma_v$) as they are narrower than the detector sensitivity. All the best fit of the features are provided in Table~\ref{gauss_rdpha} and they are all statistically significant for the corresponding spectrum. An absorption line at $\sim2.2-2.3$ keV is found in both spectra that we identify as the residuals of calibration around the instrumental edge of Au-M at 2.3 keV. We also highlight that the residuals associated to these features are clearly stronger in the RDCTI data (more than 10$\sigma$) than in the RDPHA spectrum (less than 5$\sigma$), suggesting that RDPHA calibrations provide a better correction at low energies.
Other marginally statistically acceptable features may also be found in the RGS spectra, but they are not taken into account as they are beyond the scope of this paper. In addition, an absorption edge at $\sim8-9$ keV (associated to highly ionised species of Iron, Fe XXI - XXV) is observed and it was then included in the fit. 

The absorption lines of the RDPHA corrected spectrum, ordered by energy as shown in Table~\ref{gauss_rdpha}, can be associated to Fe XXV $K_{\alpha}$ (6.70 keV), Fe XXVI $K_{\alpha}$ (6.99 keV), Fe XXV $K_{\beta}$ (7.86 keV) and Fe XXVI K$_\beta$ (8.19 keV), respectively. Notably, the centroid energy of these features are only marginally affected by the continuum and are compatible with zero shift although the uncertainty on them is of the order of $\sim900$ km s$^{-1}$. In addition, we note marginal hints of the K$_\alpha$ lines of S XVI $K_{\alpha}$ (2.64 KeV), Ar XVIII $K_{\alpha}$ (3.30 keV) and Ca XX $K_{\alpha}$ (4.10 keV), but they are not statistically significant.

On the other hand, the RDCTI corrected spectrum shows a number of features whose energies are not consistent with those found in the RDPHA corrected spectrum. Indeed, we detected absorption lines at 6.55 keV, 6.83 keV, 7.69 keV and 7.99 keV (see Table~\ref{gauss_rdpha}). 
In Figure~\ref{comparison_rdpha_rdcti}, we show the significant discrepancies of the centroid energies of absorption lines in the RDCTI and RDPHA corrected spectra. 
Hence, those found in the RDCTI data could be either different line species or miscalibrations of the energy scale with one of the two corrections. We add that the energy lines in the RDCTI spectrum do not appear to be consistent with known rest frame absorption lines: we might claim that the highest energy line (7.99 keV) can be associated to the same Fe XXVI K$_{\beta}$ line of the RDPHA corrected spectrum, but with a high redshift ($\sim 9500$ km s$^{-1}$). Adopting a similar argument also to the lines at 6.55 keV and 6.83 keV, and associating them to Fe XXV and Fe XXVI, we would expect a redshift of $\sim6000/7000$ km s$^{-1}$. On the other hand, a similar approach can be applied to the line at 7.69 keV that, if associated to Fe XXV $K_{\beta}$, would be redshifted of $\sim 4000$ km s$^{-1}$. Alternatively, the lines at 6.83 keV and 7.69 keV might be associated to blueshifted lines of Fe XXV and Fe XXVI with velocity of $\sim6000$ and $\sim30000$ km s$^{-1}$, respectively, but they are larger than the velocities commonly observed in the dippers.

However, these claims are only qualitative and might be misleading. Hence, in order to better constrain the properties of the warm medium which is more likely responsible for the narrow absorption lines, we substitute the Gaussian models with an {\sc xstar} grid \citep{kallman01}. The selected {\sc xstar} grid depends on the column density of the warm absorber, its ionisation level and its redshift/blueshift velocity. The dispersion velocity is not a variable parameter of the grid and is fixed to zero. In Table~\ref{gauss_rdpha}, we show that for the RDPHA spectrum the warm absorber column density is about an order of magnitude higher than that of the RDCTI spectrum ($60\times10^{22}$ cm$^{-2}$ vs $4\times10^{22}$ cm$^{-2}$). The ionisation is also clearly higher for the RDPHA spectrum ($\sim4.2$ vs $\sim3.4$). 

This latter result is consistent with the energy of the Iron edges found in both spectra.
Indeed, the edge in the RDCTI corrected spectrum is at 8.4 keV, which energy may be associated to a lower level of Iron ionisation (Fe XXIII-XXIV) if compared to the RDPHA corrected one ($\sim8.8$ keV, Fe XXV). In addition, we note that if the RDCTI edge is associated anyway with the Fe XXVI or XXV K-edge, we should consider a high redshift ($>15000$ km s$^{-1}$). On the other hand, this redshift is not consistent with the blueshift of $2600$ km s$^{-1}$ found with the {\sc xstar} model (Table~\ref{gauss_rdpha}), unless to hypothesize that the edge is produced near the compact object and is affected by relativistic redshift. We finally mention that {\sc xstar} leaves stronger residuals around 6.5-7.2 keV in the RDCTI data, suggesting that, in this spectrum, either the lines are broad (as the dispersion velocity is fixed at zero) or {\sc xstar} is not able to simultaneously well model all the lines present in the spectrum.

The {\sc xstar} grid in the RDPHA spectrum provides instead a blueshift of $\sim 300$ km s$^{-1}$ which better matches the blueshift of the absorption lines found in other \textit{XMM-Newton} and \textit{Chandra} observations of GX 13+1 (we will discuss a direct comparison with the \textit{Chandra} data in Section~\ref{xmm-chandra}). 

\subsection{Broad emission line}
\label{emissionline}
The comparisons in the previous section can be further extended investigating the properties of the broad Iron emission line. We initially model it with a Gaussian component in which we left the $\sigma_v$ parameter free to vary, as the line is clearly broad. However, we limited its range to 0.7 keV, in order to avoid unphysical values. 

\begin{figure}
\center
\hspace{-0.5cm}\includegraphics[height=8.5cm,width=6.5cm,angle=270]{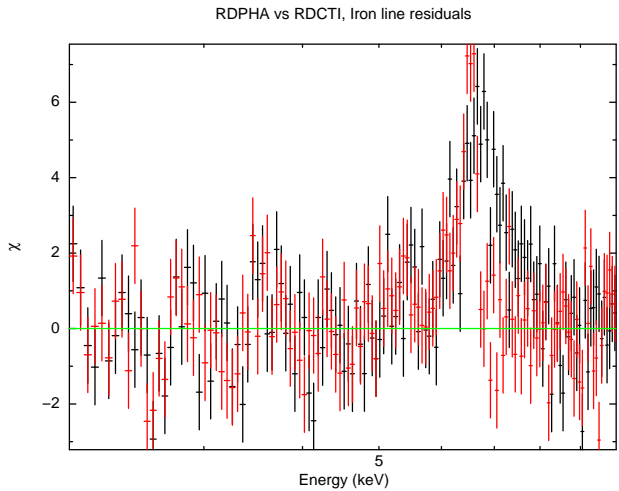}
\caption{Residuals of the best fit continuum model and absorption features, in the 2.0-10 keV energy range, for the RDPHA (\textit{black}) and RDCTI (\textit{red}) spectra. As expected, an emission line is clearly found at $\sim6-7$ keV and it displays a different shape for the two calibrations (see text).}
\label{RDPHA_RDCTI_diskline_zero}
\end{figure}

In Figure~\ref{RDPHA_RDCTI_diskline_zero}, we show the residuals of the best fit continuum model and absorption features: a clear emission feature is seen, as expected, at $\sim6-7$ keV and we note that the spectral shape is different adopting the two calibrations.
This feature seems to suggest that the RDCTI correction produces a marginally less physical energy of the Iron line (6.3 keV), which is expected to be observed between 6.4 and 7.0 keV (depending on the iron ionisation level), although the error bars make the feature being consistent with 6.4 keV, i.e. neutral Fe. We found that its broadness is $\sim0.3$ keV. On the other hand, the RDPHA corrections provide an energy line of 6.6 keV which is (well) consistent with the energies found in \citet{diaz12} and \citet{dai14}, and it can be associated to Fe XXV. However, its broadness is larger (0.7 keV) than that of the RDCTI corrected spectrum. We note that also its intensity is about a factor of 4 stronger than in the RDCTI data.
The discrepancy in the two spectra for the observed properties of the broad Gaussian profile could be due either directly to the diverse calibration of the energy scale or to the difference in the underlying modeling of the continuum or also to a mismodeling of the line itself.

\begin{figure*}
\center
\subfigure{\includegraphics[height=7.8cm,width=6.9cm,angle=270]{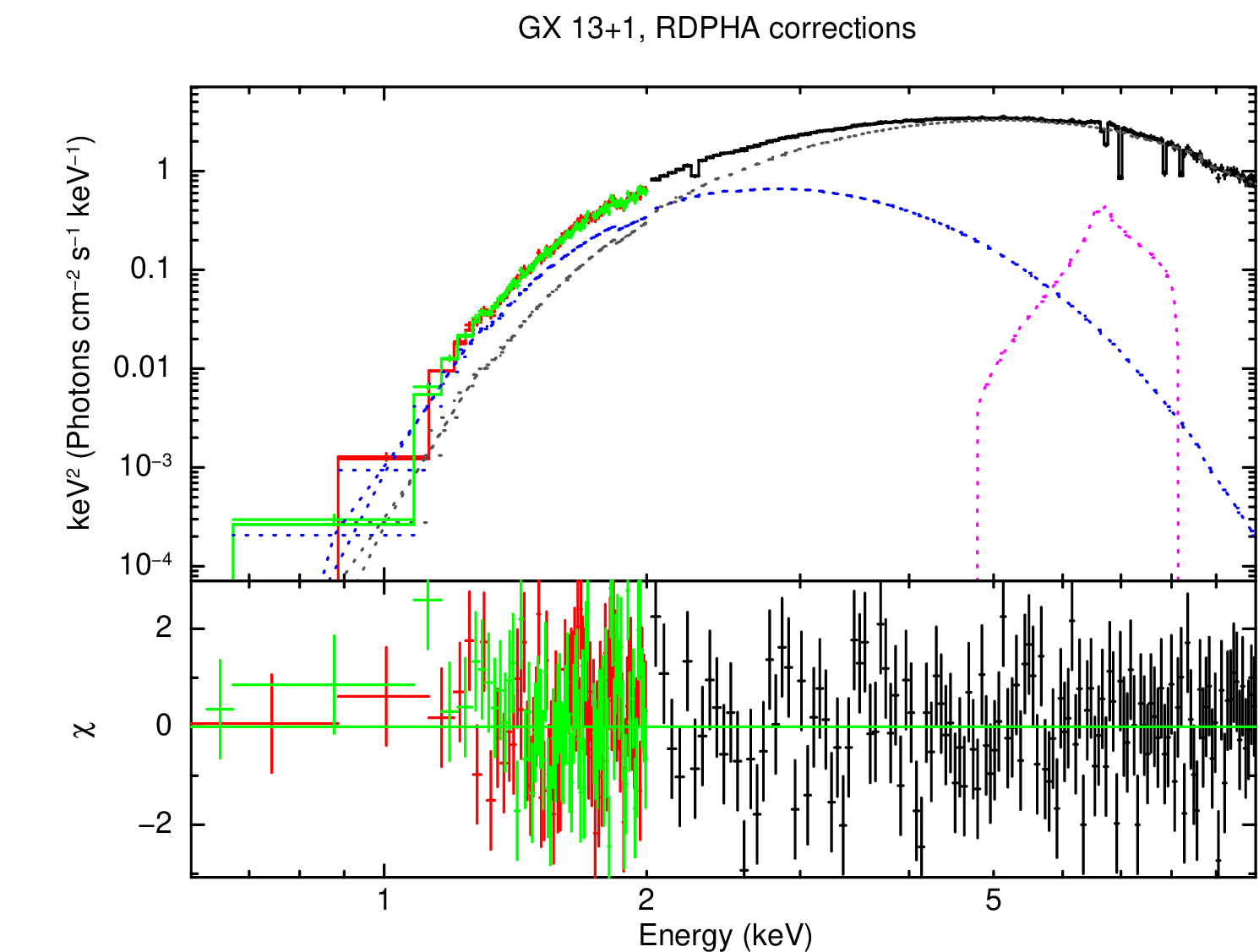}}
\subfigure{\includegraphics[height=7.8cm,width=6.9cm,angle=270]{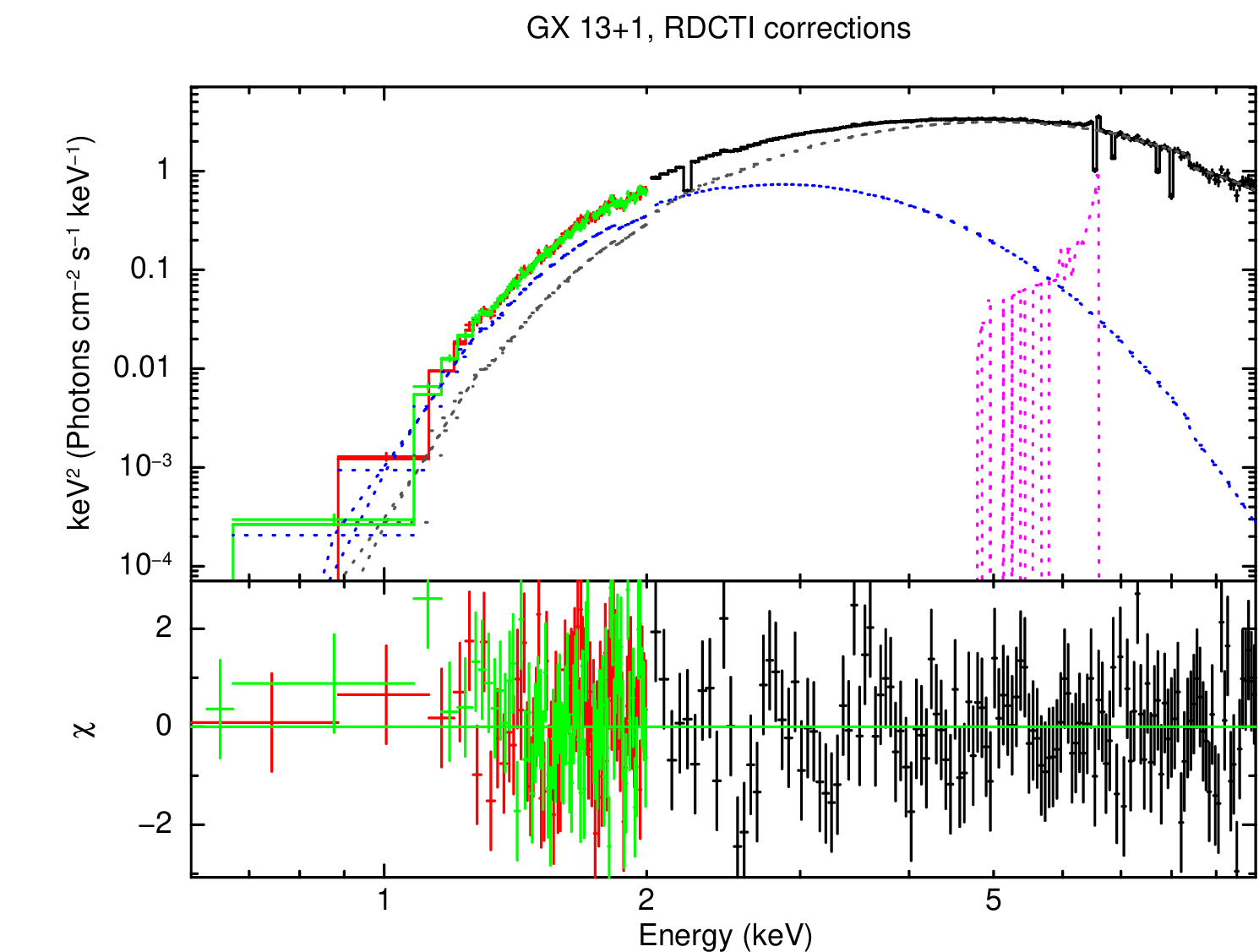}}
\caption{Unfolded $Ef(E)$ EPIC-pn (\textit{black}) and RGS spectra (\textit{red} and \textit{green}), corrected with RDPHA (\textit{left}) and RDCTI (\textit{right}). The solid line represents the best fit model, the dashed \textit{blue} curve is the {\sc bbody} component, the dashed \textit{gray} curve is the {\sc nthcomp} component and the dashed \textit{purple} curve is the {\sc diskline} component. A number of absorption features is taken into account in the fit (see text). For display purposes, the RGS spectra have been rebinned at a minimum significance of 15$\sigma$.}
\label{RDPHA_continuum+gauss-reflionx}
\end{figure*}

However, as we have already shown that the continuum parameters are insensitive to the detailed calibration of the energy scale and the width of the line suggests also relativistic smearing, we substitute the Gaussian model with a {\sc diskline} model \citep{fabian89} in order to describe a relativistic reflection line. This model depends on six parameters: the energy of the line, the radius of the inner and outer disc ($R_{in}$ and $R_{out}$) in unit of gravitational radius, the inclination angle of the system, the power-law index (\textit{Betor10}) in the radial dependence of the emissivity, and finally the normalization.
We do not allow the energy line to go beyond the range 6.4-7.0 keV as it represents the lower and upper energy limits of the K$_{\alpha}$ Iron emission lines in all possible ionisation states. 

In Fig.~\ref{RDPHA_continuum+gauss-reflionx}, we show the unfolded spectra related to the best fits model. Although the shape of the lines appears different, in general, the continuum parameters are consistent within the errors with those inferred adopting the simple Gaussian line (see Table~\ref{table_continuum_gauss_rdpha}). 
The emission energy line in both RDCTI and RDPHA data are consistent with 6.6-6.7 keV (Fe XXV), while the emissivity index is constrained between -2.4 and -2.6. In addition, although the inferred inner disc radius ($R_{in}$) is poorly constrained, our results point towards $R_{in}\sim$ 6 R$_{g}$ for both the RDPHA and RDCTI corrected spectra, while the outer radius has been fixed to $10^4$ R$_g$ as it was unconstrained.
The inclination angle is instead extremely different for the two corrections as a large inclination angle ($>50$\textdegree) is found for the RDPHA corrected spectrum while, for the RDCTI corrected spectrum, the inclination angle is consistent with a value lower than $\sim30$\textdegree. The latter result collides with the findings of dips in the lightcurves of GX 13+1, which are usually observed in sources with high inclination angles ($>65$\textdegree).

\begin{figure*}
\center
\subfigure{\includegraphics[height=7.8cm,width=6.9cm,angle=270]{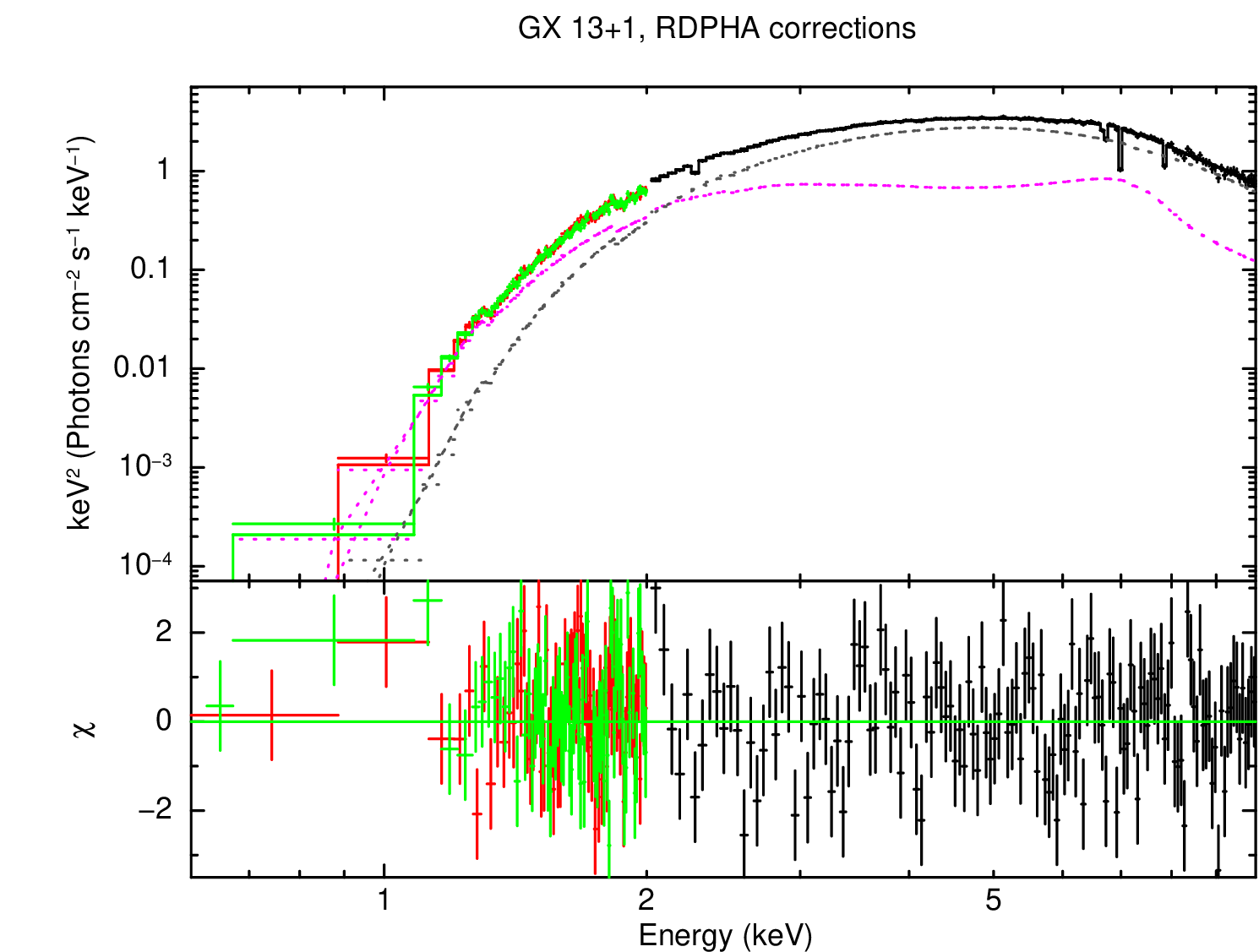}}
\subfigure{\includegraphics[height=7.8cm,width=6.9cm,angle=270]{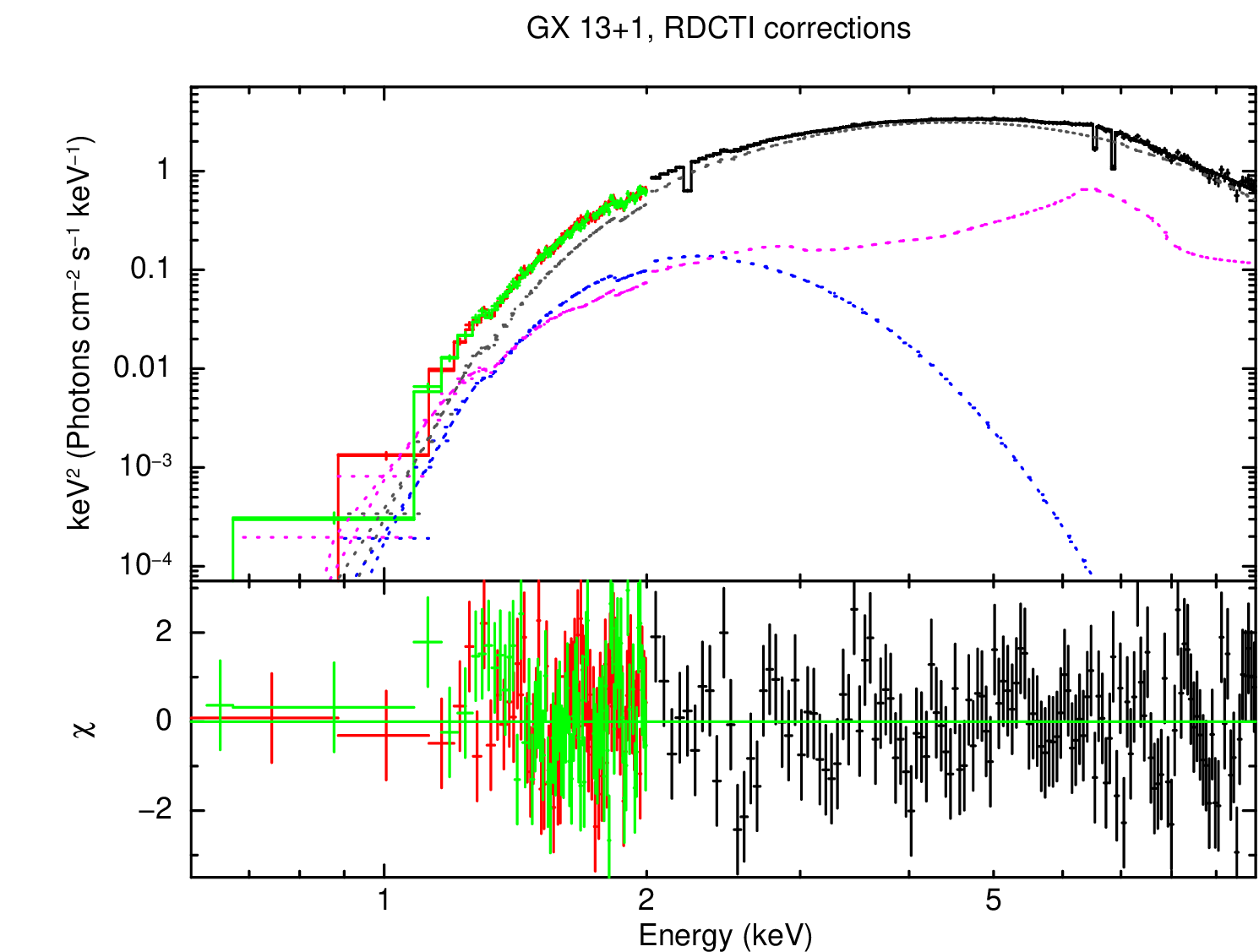}}
\caption{Unfolded $Ef(E)$ EPIC-pn (\textit{black}) and RGS spectra (\textit{red} and \textit{green}), corrected with RDPHA (\textit{left}) and RDCTI (\textit{right}). The solid line represents the best fit model, the dashed \textit{blue} curve is the {\sc bbody} component, the dashed \textit{gray} curve is the {\sc nthcomp} component and the dashed \textit{purple} curve is the {\sc reflionx} component. A number of absorption features is taken into account in the fit (see text). For display purposes, the RGS spectra have been rebinned at a minimum significance of 15$\sigma$.}
\label{RDPHA_continuum+gauss}
\end{figure*}

\subsection{Reflection component}

However, the {\sc diskline} model gives account of the shape of a single emission line and does not describe the whole reflection emission and this may lead to a wrong estimate of the inclination angle, for example. Hence, we refined our previous results substituting the {\sc diskline} model with a full broadband self-consistent reflection model, i.e. the {\sc reflionx} model \citep{ross05}. It takes into account the reflection continuum and a set of discrete features. The reflection component close to the NS should be affected by Doppler and relativistic effects in the inner regions close to the compact object which are not included in the model. Hence, we multiplied the reflection component by the relativistic kernel {\sc rdblur} that depends on the inner disc radius, the emissivity index ($Betor10$), the inclination angle and the outer disc radius. The latter has again been fixed to $10^4$ R$_g$, as it turned out to be unconstrained. In addition, we also introduced an {\sc highecut} component which allowed us to physically constrain the highest energy range of the reflected emission. We fixed the low energy cut-off at 0.1 keV, while the folding energy cut-off was tied to the electron temperature of the comptonising component as $2.7\times$kT$_e$ since, for saturated comptonisation, a Wien bump is formed at $\sim 3$ times the electron temperature. Finally, we linked the photon index of the {\sc nthcomp} model to that of the reflection component. 

In Figure~\ref{RDPHA_continuum+gauss} and Table~\ref{table_continuum_gauss_rdpha} (columns $\#7$ and 8), we show the best fit parameters for the RDPHA and RDCTI corrected spectra. 
The addition of the broad-band reflection component modifies the spectral description of the continuum, i.e. the parameters of the blackbody and the Comptonized components. Not surprisingly the photon index of the {\sc nthcomp} model pegged (or is very close) to 1.4, since the {\sc reflionx} model is not calculated for $\Gamma$ below 1.4. However, in the previous section, we found that the photon index of the {\sc nthcomp} would prefer to settle close to 1. Therefore, it is important to mention that the use of {\sc reflionx} might force the fit to converge towards spectral parameters of the broadband continuum which are affected by this assumption.

We note that in the case of the RDPHA corrected spectrum, the {\sc nthcomp} component dominates in whole bandpass, with a blackbody emission stronger than the reflection one at energies below $\sim4$ keV. However, for this best fit, the ionisation parameter of the {\sc reflionx} is dramatically low ($\sim15$ erg cm s$^{-1}$) and, moreover, the normalization of the soft component is mostly unconstrained. This may suggest that the properties of the spectrum does not allow us to describe the overall continuum with both the {\sc reflionx} and the {\sc bbody} components. Therefore, in Table~\ref{table_continuum_gauss_rdpha}, we show also an alternative fit without the {\sc bbody} component whose spectral parameters appear instead physically more plausible. Indeed the ionisation parameter is now increased up to $\sim1100$ erg cm s$^{-1}$, which is more acceptable than the previous $\sim15$ erg cm s$^{-1}$. Hereafter, we consider this fit as reference for the RDPHA data. 

On the other hand, the RDCTI corrected spectrum does not suffer of this degeneracy in the spectral parameters. The {\sc bbody} component is well constrained, with the reflection component that dominates over the blackbody emission and is predominant at energies below 1.5 keV. The parameters of ionisation is consistent with $\sim200 $ erg cm s$^{-1}$. 
The discrepancies in the reflection properties inferred with {\sc reflionx} between RDPHA and RDCTI corrected spectra again suggest that the shape of the Iron emission line may be different for the two spectra as found with a simple {\sc gaussian} or {\sc diskline} model.

Then, we further note that the column density of the RDCTI has raised up to $3.4\times10^{22}$ cm$^{-2}$ while that of the RDPHA (in the fit without a blackbody) settles at $\sim3.7\times10^{22}$ cm$^{-2}$. In addition, the abundance of iron relative to solar value is over-abundant ($\sim3$, which was set as upper limit in order to avoid unphysical values). The latter result is found also in the RDPHA data, although the error bars in the parameters of both spectra are large.  

Finally, the {\sc RDBLUR} parameters that account for the relativistic smearing of the reflection component give similar best fit parameters for the RDPHA and RDCTI data. In particular the inner disc radius is consistent with $\sim 6-15$ R$_g$ and the inclination angle is larger than $\sim50$\textdegree, which would be consistent with the expected high inclination due to the presence of dips. 

The spectral parameters obtained from the fits of the broad component clearly suggest that RDCTI and RDPHA corrections provide different spectral shapes. However, they do not allow us to univocally discriminate between the goodness of the two corrections. We can only note that different spectral results are obtained in the two cases.

\subsection{Comparing \textit{XMM-Newton} and \textit{Chandra} data}
\label{xmm-chandra}

\begin{figure}
\center
\includegraphics[height=8.5cm,width=7.5cm,angle=270]{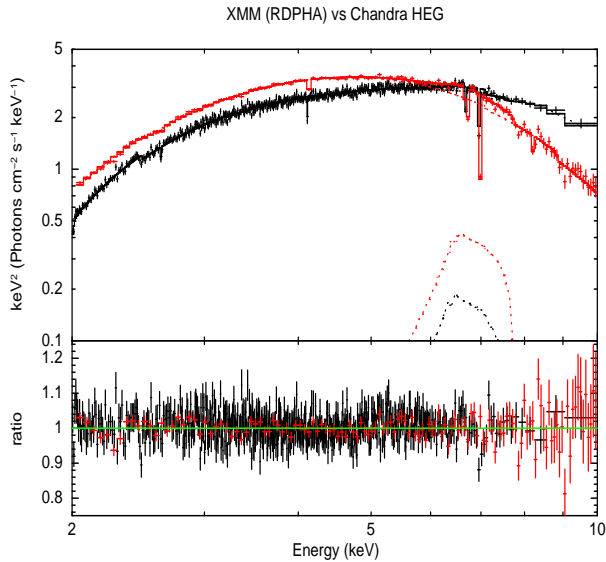}
\caption{\textit{Top-panel}: Unfolded $Ef(E)$ \textit{Chandra}-HEG (\textit{black}) and EPIC-pn RDPHA corrected spectra (\textit{red}), adopting an absorbed {\sc edge$\cdot$(nthcomp+diskline+gauss)} (see text). The positions of the absorption features which are common to the two spectra are widely consistent within the errors, when fitted simultaneously. \textit{Bottom panel}: Ratio of the data and the best fit model.
For display purposes, the HEG spectrum have been rebinned at a minimum significance of 30$\sigma$.}
\label{RDPHA_HEG}
\end{figure}

We found that the RDPHA and RDCTI corrections provide similar broadband continua and the study of the broad iron emission line suggests that also the inclination angle is compatible with that inferred by the existence of dips in the lightcurve. However, we note that the most important discrepancy (beyond the ionisation level of the reflection component) between the two corrections turns out to be the centroid energy line of the absorption features which significantly differ for the two corrective approaches of spectra. However, the RDPHA corrected spectrum shows absorption features whose energies are largely consistent with those found in the Chandra observation presented in \citet{dai14}. To better assess this issue, we fitted simultaneously the HEG and RDPHA spectra, adopting a continuum model consisting of an absorbed {\sc nthcomp} and a {\sc diskline}. We added Gaussian models, with dispersion velocity fixed at 0, in order to fit the lines at 6.69 keV, 6.70 keV and 8.2 keV (the line at $\sim 7.8$ keV is not present in the \textit{Chandra} spectrum), plus the addition of an absorption edge at $\sim8.8$ keV. These features are the focus of such a spectral comparison as they are common to the \textit{Chandra} and EPIC-pn RDPHA corrected spectra.
Furthermore, in HEG data, we also considered significant lines of Ca XX K$_{\alpha}$, Si XIV K$_{\alpha}$ and Si XVI K$_{\alpha}$ lines at 4.10 keV, $\sim 2.0$ keV and $\sim 2.6$ keV respectively \citep{dai14}, which are not (or, at maximum, marginally) seen in the EPIC-pn data.
We left the continuum free to vary between the spectra because the HEG data can show spectral variability and they are also affected by pile-up. Since this is not taken into account, the continuum spectral shape can be different from that of the RDPHA spectrum. This is supposed not to be an issue for the absorption features as we checked that they are largely independent from the continuum model and are usually found also at different levels of luminosity \citep{ueda04}. On the other hand, the energy lines are linked between the spectra during the fit calculation, except their normalisations. In Figure~\ref{RDPHA_HEG}, we show the best fit obtained with the mentioned model. The best fit energy lines of the common lines are 6.702 ($\pm0.008$) keV, 6.98 ($\pm0.01$) keV and 8.19 ($\pm0.09$) keV, and the edge is at 8.84 ($\pm0.08$) keV which are widely consistent with those found in Section~\ref{absfeatures}.

It is evident that no significant residuals are still present except for a HEG point at $\sim 7$ keV, which present a residual at $\sim4\sigma$, and the EPIC-pn absorption lines at $\sim7.9$ keV and $\sim2.3$ keV. We suppose that the former is produced by either a broadening of the line in the \textit{Chandra} data or to a blueshift of the line which cannot be detected in the EPIC-pn spectrum; on the other hand, the $\sim7.9$ keV line is the Fe XXV K$_{\beta}$ which is not observed in the HEG data while the 2.3 keV line is instead likely a residual in the calibration of the EPIC instrument around the Au edge. This may suggest that the RDPHA calibrations may be further improved.
In any case, as the other absorption features are totally consistent between HEG and EPIC-pn data, we strongly suggest that the RDPHA calibrations should to be preferred to the RDCTI ones.

\section{Discussion} 

In this work we have analysed a single \textit{XMM-Newton} observation taken in \textit{Timing} mode of the LMXB dipping source GX 13+1.
The main goal of this paper is studying the impact of different calibrations (RDPHA and RDCTI) of the energy scale in EPIC-pn Timing Mode on the spectral modeling of the LMXB dipping source GX 13+1.
The RDPHA is an empirical correction of the energy scale that does not assume any specific energy-dependence, as it is instead for the RDCTI correction.
We have proven that RDPHA and RDCTI corrected data provide different spectral results and we tentatively try to understand which correction offers the most plausible and physically acceptable scenario. We have shown that, in order to avoid spurious effects on the spectral analysis due to the existence of pile-up, it is necessary to remove the three brightest, central columns of the EPIC-pn CCD.

We found that the broad band continuum can be well described for both types of spectral data by the combination of a soft blackbody and an optically thick, cold comptonising component. This result is consistent with that found for other accreting NSs and also for \textit{XMM-Newton} and \textit{Chandra} spectra of GX 13+1 (e.g. \citealt{diaz12,dai14} and reference therein). The soft component provides an inner temperature consistent with $\sim 0.6$ keV for the two corrections. From its normalization, we infer an emission radius of $\sim40$ km, which prevents to relate this emission with the NS surface and allows us to more likely associate it to the accretion disc.
The parameters of the comptonising corona, which may be possibly produced close the NS, are instead consistent with a cold ($\sim1.2$ keV) electron population, where we assumed that the seed photons are provided by the inner disc regions for both spectra. 

Notably, several features are clearly observed in the RDPHA and RDCTI corrected spectra and it has been suggested that they are produced by a warm absorber during both dips and persistent epochs of GX 13+1. The warm absorber may be created by outflows at the outer regions of the disc where the thermal pressure is stronger than the relative gravitational pull (e.g. \citealt{diaz12,dai14}). 
The narrow absorption features are the best gauge to estimate the accuracy of the energy scale yielded by the two aforementioned calibration methods.
We compared our absorption lines, modeled with a Gaussian, to the absorption lines observed in the \textit{Chandra/HEG} data presented in \citet{dai14}. In that work, the authors found the existence of lines at 2.6234, 4.118, 6.706, 6.978, 8.273 keV associated with S XVI, Ca XX, Fe XXV, Fe XXVI K$_{\alpha}$ and Fe XXVI K$_{\beta}$, respectively, with possibly blueshifts comprise between $\sim200-1000$ km s$^{-1}$. We detected the same lines (although the first two are not statistically significant) only on the RDPHA data, with the addition of a line at 7.82 keV, more likely associated to Fe XXV K$_{\beta}$. Unfortunately the error bars inferred by a simple Gaussian model are large and prevents to constrain the possible blueshift of most of these features. This may be done only to the line of Fe XXVI K$_{\alpha}$ which is shifted if compared to the rest frame energy (6.9662 keV), suggesting a blueshift of $\sim 1500\pm 300$ km s$^{-1}$. 
On the other hand, the lines in the RDCTI data are systematically different and shifted from those found in the RDPHA corrected spectrum. These absorption features, if associated to the lines observed in \textit{Chandra} data, should have too large redshifts ($>5000$ km s$^{-1}$; see section~\ref{absfeatures}) which are physically implausible if produced by a warm absorber located at the outer edge of the disc. 

We note that, however, these associations might then be misleading. 
Therefore, we tentatively tried to better constrain the properties of the warm medium which produces these absorption lines adopting an {\sc xstar} grid. It showed us that for the RDPHA spectrum, the column density of the ionised medium is $\sim6\times10^{23}$ cm$^{-2}$, with an ionisation level of Log($\xi$) $\sim 4.2$, and more likely ejected by the system with a velocity of $\sim 300$ km s$^{-1}$. On the other hand, for the RDCTI data, column density and ionisation are an order of magnitudine lower while the blueshift velocity is instead a factor of 8-9 higher than the RDPHA data. However, the blueshift velocity in the RDPHA corrected spectrum is highly consistent with those found in previous works \citep{ueda04,madej13,diaz12,dai14}. In addition, the {\sc xstar} grid is not completely able to model the lines at 6-7 keV in the RDCTI spectrum, suggesting either a broadening or a general mis-modeling of the lines.
This result further supports the conclusion that the RDPHA corrections are generally more reliable than the RDCTI ones. 

We then found that RDCTI and RDPHA corrections provide similar results on the continuum and the broad emission line, if the latter is described by a model more complex than a simple Gaussian.
In fact, we found the existence of residuals which suggest a relativistic broadening of the line. For this reason, we initially introduced the {\sc diskline} model which shows that the inclination angle is small ($<30$\textdegree) for the RDCTI data and not consistent with the dip episodes of GX 13+1, which instead point towards a large inclination angle ($60-85$\textdegree). 
On the other hand, this finding is, however, not confirmed if we model the disc reflection with the self-consistent disc reflection code {\sc reflionx} \citep{ross05} modified by a relativistic kernel. Indeed, for both spectra, we found that the inclination angle can be larger than {50\textdegree}, confirming that a single Gaussian or {\sc diskline} model are too simple for the quality of the data. 
However, we highlight that the results obtained with {\sc reflionx} can be affected by the pitfall of the model, as {\sc reflionx} is not defined for photon index lower than 1.4 while we found that both type of spectra can be fitted by an {\sc nthcomp} with $\Gamma\sim1.0$. In addition, {\sc reflionx} does not take into account the self-ionisation of the accretion disc, introducing a possible source of uncertainty in the description of the continuum. 
With that in mind, for the fit of the RDPHA spectrum, we observed a degeneracy in the spectral parameters of the {\sc bbody} and {\sc reflionx} component, or in other words, we could find two best fits which are statistically comparable: in the first one, the normalization of the soft component is poorly constrained and the ionisation parameter of the {\sc reflionx} is close to its lower limit ($\sim 15$ erg cm s$^{-1}$); in the other case, the soft component can be removed from the fit and the ionisation parameter converges towards more physical values ($\sim 1000$ erg cm s$^{-1}$). The spectral degeneracy warns that the data are possibly not adequate to constrain the properties of the blackbody emission, when using {\sc reflionx} because of the complexity of the adopted model. We highlight that in the first fit, the reflection systematically needs to converge towards a strong iron emission line in low ionisation conditions. This is in contrast with the existence of H-like and He-like Iron absorption features that can exist only in case of ionisation higher than 100 erg cm s$^{-1}$ \citep{kallman04}. On the other hand, this is instead satisfied in the second best fit, where the ionisation parameter is higher than 1000 erg cm s$^{-1}$. The discrepancy with the latter ionisation value and that found with the {\sc xstar} grid may more likely suggest that the density of the material is different in the warm absorber and in the disc.  
Instead, for the RDCTI data, the ionisation parameter of the reflection component is lower ($\sim200$ erg cm s$^{-1}$) than the second best fit of the RDPHA data and only marginally supports the value found with the {\sc xstar} grid ($>1000$ erg cm s$^{-1}$). Such a low ionisation level of the reflection component may collide with the energies of the absorption lines that, according to {\sc xstar}, would be expected at energies higher than those found in RDCTI corrected spectrum. 
However, as the ionisation parameter depends on the density, we note again that the absorption features are more likely produced in a warm medium. This should have a density ($10^{22}$ cm$^{-2}$) lower than that in the accretion disc, where the reflection is produced. However, as the ionisation parameter depends also on the distance, we cannot exclude the effect of the latter on the estimates of the ionisation.

We finally conclude that the reflection component cannot be easily investigated to infer the goodness of one correction in comparison to the other. Instead, it clearly suggests that the spectra, i.e. the shapes of the broad emission line, obtained with the two calibrations are different. 
However, the quality of the two corrections can be discriminated studying the narrow absorption lines. Indeed, adopting simple continuum models and Gaussian models for the narrow absorption features, the RDPHA corrected spectrum of GX 13+1 provides more physical spectral parameters. In particular, regarding the absorption lines that are much more consistent with those inferred also by \textit{Chandra}. In addition, the residuals at the energy of the instrumental Au edge (2.2-2.3 keV) are smaller in the RDPHA data, favouring a better calibration of that energy range with the RDPHA corrections. For these reasons, although the EPIC-pn calibrations can be further improved, we propose that the RDPHA corrections should be generally preferred to the standard RDCTI (or \textit{epfast}) corrections, especially in case of spectra with a large number of absorption features. Hence, supported also by our results, RDPHA will be the default calibration of the next SAS version (SAS v.14), superseding RDCTI (i.e. \textit{epfast}) that was the default from SAS v.9 to SAS v.13.5.
Our conclusions will be further tested on other accreting sources in order to support with more solid basis the RDPHA corrections.

\section{Conclusions}

Accuracy of the scale energy calibration in \textit{XMM-Newton} data taken in Timing mode is extremely important when observing bright sources, where rate-dependent effects have to be taken into account. For this reason, two calibration approaches have been developed: the new RDPHA and the standard RDCTI (\textit{epfast}) corrections. 

The aim of this work was to analyse and test the two calibrations on one EPIC-pn \textit{XMM-Newton} observation, taken in Timing mode, of the persistent accreting NS GX 13+1. This source is a dipper which has shown periodic dips and its spectra are characterised by a number of absorption features. It also shows an emission line associated to the Fe XXV or XXVI K$_{\alpha}$ transition. 
Hence GX 3+1 is a suitable source to infer the goodness of the two calibrations, thanks to its several absorption features, the emission line and a simple continuum.
\noindent We found that:
\begin{itemize}
\item the continuum can be well described, in both spectra, by a blackbody of $\sim0.5$ keV and a high energy comptonisation with electron temperature of $\sim1.1-1.2$ keV and photon index of $\sim 1$; 
\item however, the two calibrations provide different results of the spectral features as the absorption lines observed in the RDCTI and RDPHA spectra differ significantly. 
\item We suggest that the lines in the RDCTI spectrum could be associated to known atomic transitions only assuming implausibly high inflow and outflow velocities for this source. On the other hand, the absorption lines in the RDPHA spectrum are more consistent with those already found in previous \textit{Chandra} and \textit{XMM-Newton} observations and we easily associated them to Fe XXV $K_{\alpha}$ (6.70 keV), Fe XXVI $K_{\alpha}$ (6.99 keV), Fe XXV $K_{\beta}$ (7.86 keV) and Fe XXVI K$_\beta$ (8.19 keV). 
\item We also observed marginal differences in the shape of the broad emission line, either when fit with a {\sc diskline} or with a {\sc reflionx} model. However, such a component cannot allow us to clearly infer the validity of the two corrections because of the poor quality of the constraints on the best fit parameters. 
\item Finally, although we note that improvement can be made especially at the energy of the instrumental Au line ($\sim2.3$ keV), our results suggest that the RDPHA calibrations are more physically reliable than the RDCTI ones and, for this reason, they should be implemented as default as of SASv14 (and associated data reduction pipeline).
\end{itemize}

\section*{Acknowledgements} 

A. R. gratefully acknowledges the Sardinia Regional Government for the financial support (P. O. R. Sardegna F.S.E. Operational Programme of the Autonomous Region of Sardinia, European Social Fund 2007-2013 - Axis IV Human Resources, Objective l.3, Line of Activity l.3.1). This work was partially supported by the Regione Autonoma della Sardegna through POR-FSE Sardegna 2007-2013, L.R. 7/2007, Progetti di Ricerca di
Base e Orientata, Project N. CRP-60529, and by the INAF/PRIN 2012-6.
The High-Energy Astrophysics Group of Palermo acknowledges support from
the Fondo Finalizzato alla Ricerca (FFR) 2012/13, project N. 2012-ATE-0390,
founded by the University of Palermo.

\addcontentsline{toc}{section}{Bibliography}
\bibliographystyle{mn2e}
\bibliography{biblio}

\end{document}